# Abrupt Enhancement of Spin-Orbit Scattering Time in Ultrathin Semimetallic SrIrO$_3$ Close to the Metal-Insulator Transition

L. Zhang,[1] X. Jiang,[1] X. Xu,[1,2] X. Hong[1,2,*]

[1] *Department of Physics and Astronomy, University of Nebraska-Lincoln, Lincoln, Nebraska, 68588-0299, USA*

[2] *Nebraska Center for Materials and Nanoscience, University of Nebraska-Lincoln, Lincoln, Nebraska, 68588-0299, USA*

[*] xia.hong@unl.edu

**Abstract**

We report a magnetotransport study of spin relaxation in 1.4-21.2 nm epitaxial SrIrO$_3$ thin films coherently strained on SrTiO$_3$ substrates. Fully charge compensated semimetallic transport has been observed in SrIrO$_3$ films thicker than 1.6 nm, where the charge mobility at 10 K increases from 45 cm$^2$/Vs to 150 cm$^2$/Vs with decreasing film thickness. In the two-dimensional regime, the charge dephasing and spin-orbit scattering lengths extracted from the weak localization/anti-localization effects show power-law dependence on temperature, pointing to the important role of electron-electron interaction. The spin-orbit scattering time $\tau_{so}$ exhibits an Elliott-Yafet mechanism dominated quasi-linear dependence on the momentum relaxation time $\tau_p$. Ultrathin films approaching the critical thickness of metal-insulator transition show an abrupt enhancement in $\tau_{so}$, with the corresponding $\tau_{so}/\tau_p$ about 7.6 times of the value for thicker films. A likely origin for such unusual enhancement is the onset of strong electron correlation, which leads to charge gap formation and suppresses spin scattering.

The Ruddlesden-Popper series iridates ($Sr_{n+1}Ir_nO_{3n+1}$) exhibit highly tunable electronic and magnetic ground states due to the strong spin-orbit coupling (SOC), which competes with the electron itineracy, the on-site Coulomb energy $U$, and the crystal fields associated with the spatially extended 5$d$-orbitals.[1-4] As the end member of the series in the large bandwidth limit ($n = \infty$), $SrIrO_3$ (SIO) is a paramagnetic semimetal,[3-5] and has been suggested to exhibit nontrivial topological phases in the orthorhombic phase [Fig. 1(a)], such as Weyl/Dirac semimetals and topological Mott insulators.[6, 7] While the perovskite SIO is metastable in bulk,[8, 9] it has been realized in high quality epitaxial thin film form,[4, 10] where a range of interesting phenomena have been observed, including strain tunable Dirac node[11-13] and metal-insulator transition (MIT),[14-16] and thickness-driven dimensionality crossover.[17, 18] Due to the highly tunable electronic state, intrinsically large SOC, and emerging interfacial Dzyaloshinskii-Moriya interaction,[19] SIO thin films and heterointerfaces are promising material candidates for developing novel spin transistors[20] and skyrmion-based topological electronics.[21, 22] Despite the emerging fundamental and practical research interests, the dominant spin relaxation mechanism in ultrathin SIO films, especially in the presence of enhanced electron correlation, remains elusive to date.

In this work, we report a comprehensive magnetotransport study of spin relaxation in high quality epitaxial SIO thin films coherently strained on $SrTiO_3$ (STO) substrates. SIO films with thickness down to 2.0 nm show semimetallic transport with complete charge compensation, while the thinner films exhibit insulating behavior. From the two-dimensional (2D) weak localization (WL) and weak anti-localization (WAL) effects, we extracted the temperature and film thickness dependences of the phase coherence and spin-orbit scattering behaviors. The spin-orbit scattering time $\tau_{so}$ exhibits a quasi-linear dependence on the momentum scattering time $\tau_p$, pointing to Elliott-Yafet (EY) mechanism dominated spin relaxation. The films approaching the critical thickness of MIT show an abrupt enhancement in $\tau_{so}$, with concomitantly reduced carrier density and increased charge mobility. A likely scenario for such unusual enhancement is that spin scattering is suppressed due to an emerging charge gap.

We deposited 1.4-21.2 nm epitaxial SIO thin films on (001) STO (1.4% compressive strain) using off-axis radio frequency magnetron sputtering (see Supplementary Materials for growth conditions). Atomic force microscopy (AFM) measurements reveal smooth sample morphology for all samples, with typical surface roughness of 1-2 Å [Fig. 1(b)].[23] X-ray diffraction (XRD) spectra show (001) growth (pseudo-cubic notation) with no appreciable impurity phases [Fig. 1(c)]. The *c*-axis lattice constant is ~4.02 Å, larger than the bulk value of 3.96 Å, which is consistent with strained SIO on STO.[15] Reciprocal space mapping (RSM) shows that even the thickest film is fully strained to the substrate [Fig. 1(d)]. For magnetotransport studies, we patterned the SIO samples into Hall bar devices with the channel length of 10 to 40 μm and length/width ratio of 1:1 or 2:1. Magnetotransport measurements were performed in a Quantum Design PPMS system combined with external SR830 Lock-In and Keithley 2400 SourceMeter at an excitation current of 100 nA – 10 μA.

Figure 1(e) shows the temperature dependence of sheet resistance $R_\square(T)$ for SIO films with various thicknesses. The 2.0 nm and thicker films exhibit metallic temperature dependence ($dR/dT > 0$) over a wide temperature range, followed by an upturn below a transition temperature $T_m$ that can be described by $ln(T)$ dependence of sheet conductance $G$ [Fig. 1(e) inset]:

$$R_\square = R_0 + A \cdot T^\alpha \qquad T > T_m \qquad (1a), \text{ and}$$

$$G = G_0 + p\frac{e^2}{\pi h} ln(T/T_0) \qquad T < T_m \qquad (1b).$$

Here $R_0$ ($G_0$) is the residual sheet resistance (conductance), and $A$, $\alpha$, $p$, and $T_0$ are fitting parameters (see Supplementary Fig. S2 for fitting details). Distinct from the hexagonal SIO single crystals,[5] the high temperature $R_\square(T)$ exhibits very weak $T$-dependence, with room temperature resistivity $\rho(300\ K)$ of $0.5 - 0.9$ m$\Omega$ cm. The power exponent $\alpha$ in Eq. (1a) is about 1.1-1.2, revealing a non-Fermi-liquid behavior. The low temperature $ln(T)$ correction to the conductance [Fig. 1(f)] is the signature behavior of a 2D electron system due to WL or electron-electron interaction.[24] The thinner films show higher $T_m$, which increases from 7 K to 21 K as the film thickness reduces from 21.2 nm to 2.0 nm [Supplementary Fig. S2(a)]. The 1.6 nm film becomes insulating over the entire temperature range, and the low temperature $R_\square(T)$ can be well described by the 2D variable range hopping (VRH) model: $R_\square \propto \exp\left[\left(\frac{T_0}{T}\right)^{1/3}\right]$ [Fig. 1(g)].[25] This transition to the strongly localized behavior occurs as $R_\square$ exceeds $h/2e^2 \approx 12.9$ k$\Omega$. The 1.4 nm film, on the other hand, exhibits a crossover from thermally activated semiconducting behavior $\exp[\Delta/2k_BT]$ at high temperature to 2D VRH at about 155 K [Fig. 1(h)]. The extracted activated energy $\Delta \approx 102$ meV is consistent with previous reports for ultrathin SIO.[17] Such thickness-driven MIT in SIO has been attributed to enhanced correlation energy in the 2D limit[17] and lattice distortion imposed by the substrate symmetry.[18] The critical thickness (about 4 unit cell) in our samples is comparable with that reported for SIO films encapsulated with STO top-layers,[17] and approaches the limit for charge gap formation.[18]

Previous band structure mapping via angle-resolved photoelectron spectroscopy (ARPES) has revealed a light mass electron pocket and a heavy mass hole pocket at $E_F$ in epitaxial SIO thin films.[11, 13] The semimetallic nature of SIO is clearly manifested in the Hall effect and magnetoresistance (MR) measurements. As shown in Fig. 2(a)-(b), the magnetic field $H$-dependences of Hall resistance $R_{xy}$ and $R_\square$ can be well described by the semiclassical two-band model,[26]

$$\rho_{xy} = \frac{(n_h\mu_h^2 - n_e\mu_e^2)H + \mu_e^2\mu_h^2(n_h-n_e)H^3}{e[(n_e\mu_e+n_h\mu_h)^2+(n_h-n_e)^2(\mu_e\mu_h H)^2]} \quad \text{(2a) and}$$

$$\Delta\rho/\rho(0) = \frac{\rho(H)-\rho(0)}{\rho(0)} = \frac{(n_e\mu_e+n_h\mu_h)^2+\mu_e\mu_h(n_e\mu_e+p\mu_h)(n_h\mu_e+n_e\mu_h)H^2}{(n_e\mu_e+n_h\mu_h)^2+(n_h-n_e)^2(\mu_e\mu_h H)^2} - 1 \quad \text{(2b).}$$

Here $\rho_{xy} = R_{xy}t$ is the Hall resistivity, $t$ is the film thickness, $\rho$ is the longitudinal resistivity, $n_e$ ($n_h$) is the electron (hole) density, and $\mu_e$ ($\mu_h$) is the electron (hole) mobility. The linear $H$-dependence of $R_{xy}$ and $H^2$-dependence of $R_\square$ have been observed in films of all thicknesses, which can be satisfied for close to be fully compensated electron and hole densities, i.e., $n_e \approx n_h$, with Eq. 2 reducing to:

$$\rho_{xy} = \frac{(\mu_h-\mu_e)}{en_{e,h}(\mu_e+\mu_h)}H, \quad \text{(3a) and}$$

$$\Delta\rho/\rho(0) = \mu_e\mu_h H^2 \quad \text{(3b).}$$

Another possibility for having a linear Hall behavior is there is only a single type of charge carrier (electron). The single carrier model, however, would lead to about two orders of magnitude discrepancy between the mobility value deduced from MR, i.e., $\Delta\rho/\rho(0) = \mu^2 H^2$, and the Hall mobility, i.e., $\mu = \frac{\sigma}{ne} = \frac{|R_H|}{\rho(0)}$. Combining this fact with previous ARPES results, we can rule out the single carrier scenario.

Fitting the Hall and MR data to Eq. 2 and assuming $n = p$, we extracted the carrier density and mobility as functions of film thickness. At $T_m$, the carrier density falls in the range of $10^{19}$ and $10^{20}$ cm$^{-3}$ [Fig. 2(c)], similar to previously reported results.[15, 16] At 10 K, the 2.0 nm and 2.8 nm films show considerably lower carrier density than those at $T_m$, consistent with an emerging charge gap at the ultrathin limit.[17, 18] The carrier mobility, on the other hand, is significantly enhanced in the thinner films [Fig. 2(d)], with $\mu_e$ at 10 K increasing from 45 cm$^2$/Vs in the 21.2 nm film to 140-150 cm$^2$/Vs in the 2.0 nm films. The mobility difference for electron and hole, defined as $\beta = \frac{\mu_e - \mu_h}{\mu_e + \mu_h}$, is less than 1.5% for all samples. The mobility values are orders of magnitude higher than those observed in strongly correlated oxides in the paramagnetic metallic phase,[27] and comparable with the non-interacting electrons in thin films of high mobility perovskite semiconductor $BaSnO_3$.[28] A possible scenario is these samples are close to the fully charge compensated regime with low density of states in both the conduction and valence bands,[11] which suppresses the intra-band elastic scattering. The inter-band scattering does not play a significant role at low temperature due to the large momentum transfer required between the electron and hole bands at the Fermi level $E_F$.[11, 13] A possible origin for the film thickness dependence of mobility is the progressively reduced structural defects associated with the epitaxial strain in thinner films. As orthorhombic SIO is meta-stable in bulk, it has been shown that the film crystallinity degrades in the thicker films, where the lattice strain is released through oxygen vacancies, dislocations, and even formation of polycrystalline grains above a critical thickness (20-40 nm).[4, 10]

Figure 3(a) shows the magnetoconductance (MC) $\Delta\sigma(H) = \sigma(H)-\sigma(0)$ of the 2.8 nm SIO film at low temperatures. The sample exhibits a negative MC at low field followed by a positive MC at high field. The positive MC is widely observed in 2D electron systems due to the WL effect, where the magnetic field suppresses the constructive back scattering of electrons. The negative MC, known as WAL, originates from spin-orbit scattering, which leads to destructive interference of back-scattered electrons that can be suppressed by the magnetic field. The WAL effect has previously been used to deduce SOC induced spin splitting in semiconductor heterostructures.[29, 30] In SIO films, the transition from WAL to WL dominated MC occurs at progressively higher field at increasing temperatures, suggesting that the dephasing/scattering mechanisms contributing to these two quantum conductance correction effects have different temperature-dependences.

Given the strong SOC in SIO and considering the effect of Zeeman splitting, we employed the Maekawa-Fukuyama (MF) model to fit $\Delta\sigma(B)$:[31]

$$\frac{\Delta\sigma(H)}{\sigma_0} = \Psi\left(\frac{H}{H_i+H_{so}}\right) + \frac{1}{2\sqrt{1-\gamma^2}}\Psi\left(\frac{H}{H_i+H_{so}(1+\sqrt{1-\gamma^2})}\right) - \frac{1}{2\sqrt{1-\gamma^2}}\Psi\left(\frac{H}{H_i+H_{so}(1-\sqrt{1-\gamma^2})}\right) \quad (4).$$

Here $\Psi = \ln(x) + \psi(1/2 + 1/x)$, $\psi(x)$ is the digamma function, and $\sigma_0 = e^2/\pi h$ is the quantum conductance. The dephasing driven by inelastic scattering and spin-orbit scattering are captured in $H_i = \frac{\hbar}{4eD\tau_i}$ and $H_{so} = \frac{\hbar}{4eD\tau_{so}}$, respectively, where $D = \frac{1}{2}v_F^2\tau_p$ is the diffusion constant, $v_F$ is the Fermi velocity, $\tau_p$ is the momentum relaxation time, and $\tau_i$ is the inelastic scattering time. The parameter $\gamma = g\mu_B H/4eDH_{so}$ couples $H_{so}$ and the $g$ factor, which describes the Zeeman effect correction. At 2-5 K, the MC data can be well described by the MF model [Fig. 3(a)]. Given the sample mobility ($\mu_{e,h}$ ~140 cm$^2$/Vs at 10 K), the criterion of low magnetic field condition ($\mu_{e,h}B < 1$) is satisfied in the entire range of magnetic field investigated. The quantum corrections to the MC persists to very high field (up to 4 T), where the classical parabolic contribution is negligible.

Figure 3(a) inset plots the extracted $H_i$ and $H_{so}$ for the 2.8 nm film, which exhibit linear and quadratic $T$-dependences, respectively. Similar temperature dependences have been observed in the 2.0-5.4 nm films, while thicker films do not show prominent WL and WAL in MC above 3 K. From $H_i$ and $H_{so}$, we obtained the inelastic scattering and spin-orbit scattering lengths using $L_{i,so} = (D\tau_{i,so})^{1/2} = \left(\frac{\hbar}{4eH_{i,so}}\right)^{1/2}$. As shown in Fig. 3(b-c), both scattering lengths exhibit power-law dependences on temperature $L_{i,so} \sim T^{-\beta}$. For $L_i$, $\beta = 0.4 \pm 0.1$, close to the expected value of $\beta = 0.5$ for electron-electron scattering induced dephasing.[24] The exponent for $L_{so}$ is $\beta = 1.0 \pm 0.1$, in sharp contrast to the nearly $T$-independent $L_{so}$ observed in semiconductor quantum wells,[30] where spin-orbit scattering is related to the spin precession process induced by the intrinsic spin-splitting in the 2D band.

Figure 4(a) summarizes the $L_i$ and $L_{so}$ obtained from the 2.0-10.2 nm SIO samples at 2 K. For films thicker than 2.8 nm, neither $L_i$ nor $L_{so}$ exhibits any apparent thickness dependence. We can thus rule out the film surface/interface states as the major electron dephasing/spin scattering source. The $L_i$ value is always higher than $L_{so}$, which satisfies the strong spin-orbital scattering condition in the MF model, i.e., $\frac{\tau_i}{\tau_{so}} > 0.183$,[31] justifying the observation of WAL at low field and WL at high field.

To confirm that these samples are in the 2D diffusive region, we calculated the elastic mean free path in these samples. The mean free path can be estimated using the Drude model $L_p = v_F \tau_p$. With $v_F = \frac{\hbar k_F}{m^*}$ and $\tau_p = \mu_{e,h} m^*/e$, where $m^*$ is the effective mass and $k_F = \sqrt{2\pi n_{2D}}$ is the Fermi wave vector, $L_p$ only depends on the mobility and 2D charge density $n_{2D} = n_{e,h} t$. Given the electron and hole bands possess closely matched density and mobility values, their mean free paths are also similar. In Fig. 4(a), we used the condition of $L_p$ exceeding the film thickness to gauge if the system can be characterized as 2D. While the 10.2 nm sample sits right at the $L_p = t$ boundary, the thinner films reside well within the 2D regime, confirming the thickness-driven dimensionality crossover. Nevertheless, the condition for the WAL regime, $L_p^2 < L_{so}^2 \leq L_i^2$, is always satisfied for the electrons and holes.[31] The magnetic length at 4 T $L_B = \sqrt{\hbar/eH} \approx 13$ nm also exceeds $L_p$, confirming that the system is in the diffusive regime for the entire field range.

We consider two commonly observed spin relaxation mechanisms: the D'yakonov-Perel' (DP) type and the Elliot-Yafet type.[32] The DP mechanism describes the spin precession induced by the spin splitting of the energy level that's randomized by the elastic scattering,[32-34] and can have both the bulk (Dresselhaus) and interface (Rashba) contributions. The bulk term is important in noncentrosymmetric materials.[29] The Rashba effect is due to the inversion field at the hetero-interfaces and surfaces, as observed in semiconductor quantum wells[30] and complex oxide heterostructures,[35, 36] and is expected to be more prominent in thinner films. The EY mechanism, on the other hand, depicts the spin dephasing via momentum scattering, *e.g.,* due to impurities or electron-phonon interactions, in the presence of SOC.[32, 37, 38] Structural defects commonly occurred in epitaxial thin films, such as cation vacancies, dislocations, and grain boundaries, can all contribute to the spin scattering through the EY mechanism. One way to distinguish these two mechanisms is to examine the relation between $\tau_{so}$ and $\tau_p$.[32] In the DP mechanism, the SOC induced spin splitting acts effectively as a magnetic field that causes electron precession. The elastic scattering randomizes this process, so that the related spin dephasing time scales with $1/\tau_p$.[29] The spin-flip process in the EY mechanism, in contrast, is facilitated by momentum scattering, with the associated $\tau_{so}$ depending linearly on $\tau_p$.[39] In SIO thin films, while the WL-WAL type measurements have previously been performed to extract $\tau_{so}$,[14, 16] the relation between $\tau_{so}$ and $\tau_p$ remains elusive due to the lack of effective approaches to control the charge mobility.[11, 14-16] For example, in the study of WAL in SIO thin films grown on LaAlO$_3$ substrates, only less than 2% variation of $\tau_p$ has been reported.[14] The film thickness of our samples, on the other hand, presents a powerful control parameter to tune carrier mobility, making it feasible to identify the dominant spin relaxation mechanism in epitaxial SIO thin films.

Figure 4(b) shows $\tau_{so}$ at 2 K as a function of $\tau_p$ for the electrons. Considering that the Dirac node is lifted by the compressive strain, we assumed a parabolic conduction band with $m^* \approx m_0$.[11-13] This assumption is supported by the observed carrier density, which is significantly higher than what is expected for a linear dispersion (see detailed discussion in the Supplementary Materials).[9, 11] To avoid the complicating effect of quantum conductance correction, we used the diffusion constant at 10 K to calculate $\tau_{so}$. As all SIO films show highly consistent, very weak $T$-dependence in $\rho_{xx}$ [Fig. 1(e)], the diffusion constant exhibits little variation below 20 K [Supplementary Fig. S2(f)]. For the thicker films, $\tau_{so}$ exhibits a quasi-linear relation with $\tau_p$, increasing from 0.093±0.008 ps in the 10.2 nm film to 0.36±0.04 ps in the 3.2 nm film, which points to an EY mechanism dominated spin relaxation. The absence of Dresselhaus contribution is not surprising given that the orthorhombic structure of SIO preserves inversion symmetry. The carrier density and film thickness dependences of $H_{so}$ also rule out a dominant presence of the Rashba effect (Supplementary Fig. S3). We then consider the possible spin scattering sources in epitaxial SIO films. As $\tau_{so}$ is enhanced in the thinner films, it is natural to rule out the roughness and defect states related to the film interface/surface as the dominant spin scattering sources. A possible bulk mechanism that contributes to the lower $\tau_{so}$ in thicker films is the structural defects associated with the epitaxial strain, such as oxygen vacancies, dislocations, and polycrystalline grain formation above a critical thickness (20-40 nm), whose density can increase progressively with film thickness.[4, 10] The last mechanism is not relevant as our samples are well below the critical thickness, and XRD studies yield no sign of impurity phase growth. We also note that the charge mobility increases in thinner films, which possess lower carrier density [Figs. 2(c-d)]. This rules out a pronounced presence of charged impurities, *e.g.*, cation or oxygen vacancies, as they are highly susceptible to the screening effect. On the other hand, misfit dislocations can present even in ultrathin films,[14] and is a viable source for spin scattering.

In the EY mechanism, the momentum scattering induces spin flipping between two neighboring bands with admixed spin-up and spin-down states. The emerging energy gap $\Delta$ can suppress the spin transition probabilities, leading to enhanced $\tau_{so}$.[39] As the spin scattering rate scales with the momentum scattering rate as $\left(\frac{1}{\tau_{so}}\right)_{EY} \approx \left(\frac{L}{\Delta}\right)^2 \frac{1}{\tau_p}$, where $L$ is the SOC matrix element,[40] we deduced $L/\Delta \approx 0.55$ for the thicker films (Fig. 4b). This result is in excellent agreement with previously reported values of $\Delta \approx 28$ meV at the strain lifted Dirac point in SIO[13] and SOC induced band splitting of $L \approx 15$ meV.[11] It is important to note that the assumption of the effective mass does not change the extracted slope, as both $\tau_{so}$ and $\tau_p$ are scaled by the same $m^*$. In contrast, ARPES studies showed that the energy separation between the valence band and a neighboring band at the same momentum space is more than 0.5 eV. We thus expect the spin-orbit scattering rate associated with holes to be more than two orders of magnitude lower than that for electrons, and rule out the a notable contribution of holes in the temperature and magnetic field range investigated.

The EY-mechanism also naturally accounts for the unusually strong $T$-dependence of $\tau_{so}$ [Fig. 4(b) upper inset]. In the presence of a nearby band, spin flipping involves the electron-electron inter-band scattering,[37, 41] which can yield a $T^2$-dependent scattering rate.[42] In fact, the $T^2$-dependence of resistivity has been observed in thick SIO films on GdScO$_3$ substrates below 10 K.[15] In thin films, this effect cannot be directly observed in low temperature $R(T)$ due to the conductance correction from WL and electron-electron interaction.[24]

Close to the critical thickness of MIT, there is an abrupt enhancement in $\tau_{so}$, reaching 2.0±0.1 ps for the 2.8 nm film, and 1.9±0.1 ps and 2.2±0.1 ps for the two 2.0 nm films [Fig. 4(b) lower inset]. The two 2.0 nm films exhibit qualitatively similar magnetotransport properties (Supplementary Materials), showing such effect is robust and reproducible. For the 2.0 nm and 2.8 nm films, the $\frac{\tau_{so}}{\tau_p}$ ratio is 7.6 times of that for thicker films. There are three possible scenarios that can lead to the enhanced $\tau_{so}$ observed in ultrathin films. The first one is a sudden reduction of the defect density, which can lead to reduced momentum scattering rate $\frac{1}{\tau_p}$. There are, however, no abrupt changes observed in the mean free path $L_p$ and inelastic scattering length $L_i$ in ultrathin SrIrO$_3$ films [Fig. 4(a)], strongly suggesting that there is no sudden change in the density or type of defects in the SrIrO$_3$ films as the film thickness approaches the critical value for MIT.

The second scenario is there is a sudden reduction in $\frac{L}{\Delta}$ to 0.2. Assuming $L$ remains unchanged, we obtained an emerging energy gap of $\Delta' \approx 75$ meV. Considering the band parameters extracted from thicker SIO films ($L \approx 15$ meV, $\Delta \approx 28$ meV, $E_F > 50$ meV),[11, 13] this $\Delta'$ value is not sufficient to induce a MIT, while the corresponding $E_F$ is getting close to the conduction band edge. This is consistent with the sudden drop of the carrier density in this thickness range and the clear temperature-dependence of the carrier density, which is absent in thicker films [Fig. 2(c)]. It is also in line with the charge gap extracted in the insulating phase (~100 meV).[17] The 2.0-2.8 nm samples thus retain the metallic behavior, while the spin spin-flip rate for electron being scattered to the neighboring band is substantially reduced. With $E_F$ falling in the vicinity of the region for SOC band splitting, this scenario can also contribute to the suppressed scattering and enhanced mobility in ultrathin films [Fig. 2(d)].

We also consider the third possible scenario, where the abrupt change in $\tau_{so}$ signals a transition from EY to DP dominated spin flip scattering mechanism. Given the SrIrO$_3$ films preserves the inversion symmetry, the crossover to the DP-type can only occur when the Bloch state broadening $\Gamma \approx \frac{\hbar}{2\tau_p}$ well exceeds the energy separation from a neighboring band $\Delta$, as theoretically proposed in Ref. [40]. Using $\tau_p = 0.075$ ps as the critical value for the transition, we estimated $\Gamma$ to be about 4.4 meV. This value, however, is much smaller than the reported gap value $\Delta \approx 28$ meV for thick SIO films.[13] Therefore, the DP mechanism is not a critical spin flip mechanism in this transport regime. We thus concluded that the enhanced $\tau_{so}$ is intrinsic to ultrathin SIO due to the interplay between SOC and onset of strong correlation energy.

In summary, we report a weak (anti)localization study of spin relaxation in high quality epitaxial SrIrO$_3$ thin films. In the 2D regime, the spin-orbit scattering time in SIO exhibits quasi-linear dependence on electron mobility, pointing to the EY mechanism dominated spin relaxation, and is strongly enhanced in films close to the critical thickness of MIT. Our study reveals the complex interplay of SOC, impurity scattering, and electron correlation on spin transport in this emerging quantum material. The ultrathin films exhibit enhanced mobility and spin scattering time, making them an ideal platform for realizing novel spintronic devices such as spin transistors and SOC enhanced multiferroic tunnel junctions.

**Supplementary Materials**

See Supplementary Materials for detials of sample growth and characterization, additional magnetotransport studies, and data modeling.

**Data availability**

All relevant data that support the findings of this study are available from the corresponding authors upon request.

**Acknowledgements**

We thank Allen MacDonald and Kirill Belashchenko for valuable discussions. This work was primarily supported by the National Science Foundation (NSF) through Grant No. DMR-1710461. Additional support was provided by the Nebraska Materials Research Science and Engineering Center (MRSEC) (NSF Grant No. DMR-1420645) (materials preparation). The research was performed in part in the Nebraska Nanoscale Facility: National Nanotechnology Coordinated Infrastructure and the Nebraska Center for Materials and Nanoscience, which are supported by the National Science Foundation under Award ECCS: 1542182, and the Nebraska Research Initiative.

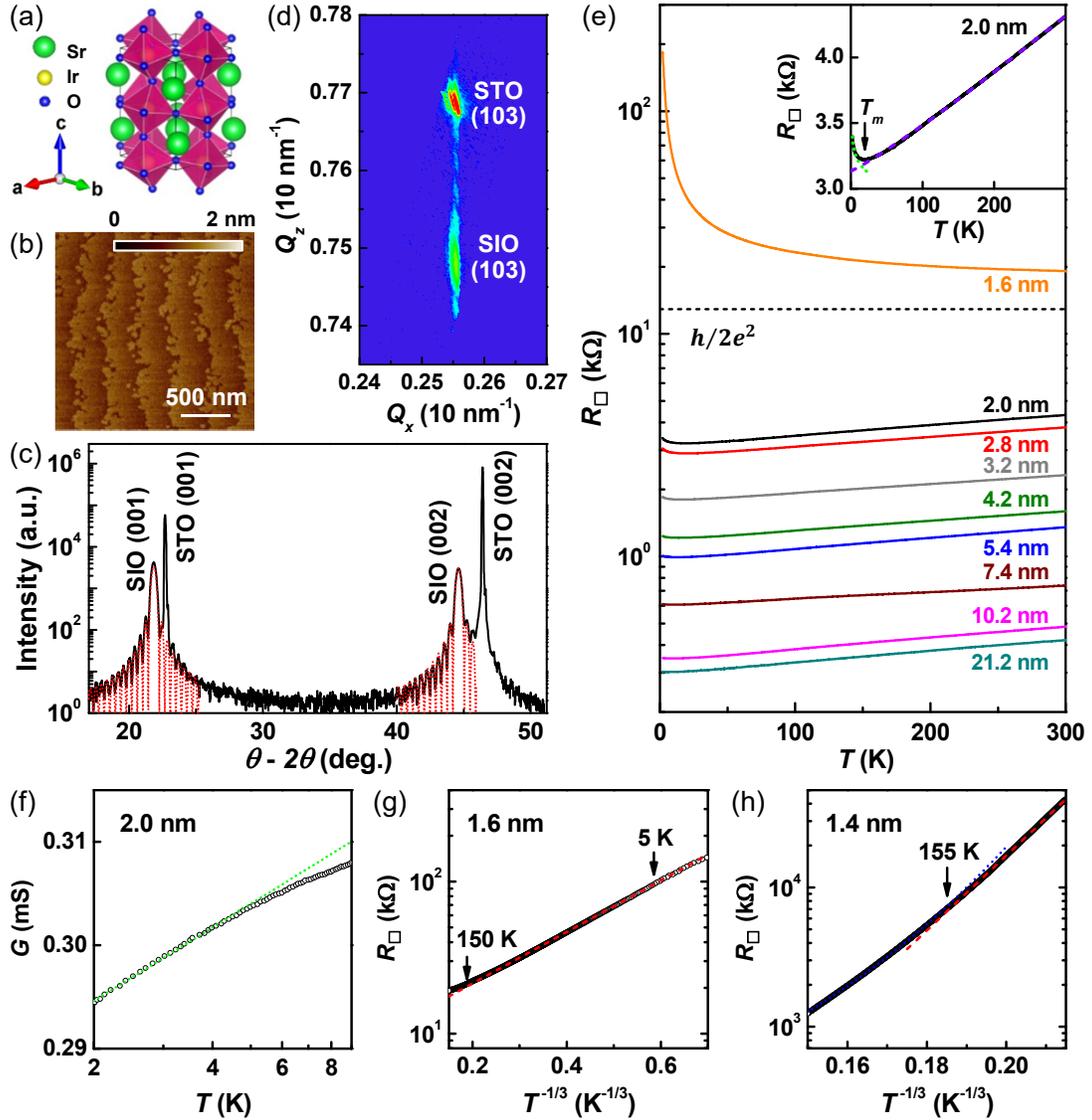

Fig. 1. (a) Schematic of orthorhombic SIO. (b)-(d) Characterizations of a 21.2 nm SIO on STO. (b) AFM topography image shows smooth film surface with flat terraces separated by 4 Å atomic steps. (c) XRD $\theta$-$2\theta$ scan with fits to the Laue oscillations around the Bragg peaks (dotted lines). (d) RSM around STO (103) peak. (e) $R_\square(T)$ for films with different thicknesses. Inset: $R_\square(T)$ for the 2.0 nm film with fits to Eq. (1a) (dashed line) and Eq. (1b) (dotted line). (f) $G(T)$ in lin-log plot for the 2.0 nm film with a fit to Eq. 1(b) (dotted line). (g) $R_\square$ vs. $T^{-1/3}$ for the 1.6 nm film with a fit to the 2D VRH model (dashed line). (h) $R_\square$ vs. $T^{-1/3}$ for the 1.4 nm film with fits to the 2D VRH model (red dashed line) and thermally activated model (blue dotted line).

Figure 2

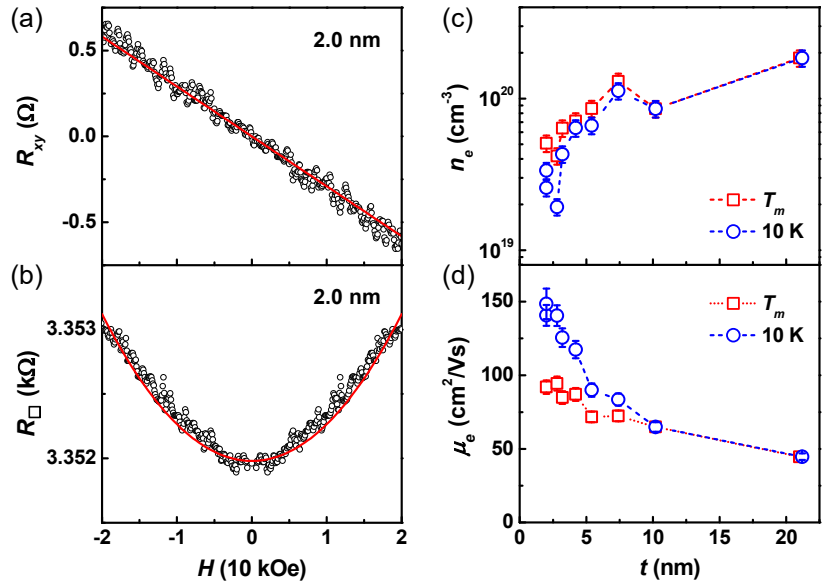

Fig. 2. (a) $R_{xy}$ and (b) $R_\square$ vs. $H$ for a 2.0 nm SIO at 20 K. The red lines are fits to Eq. 2. (c) Electron density and (d) mobility vs. film thickness at $T_m$ and 10 K.

Figure 3

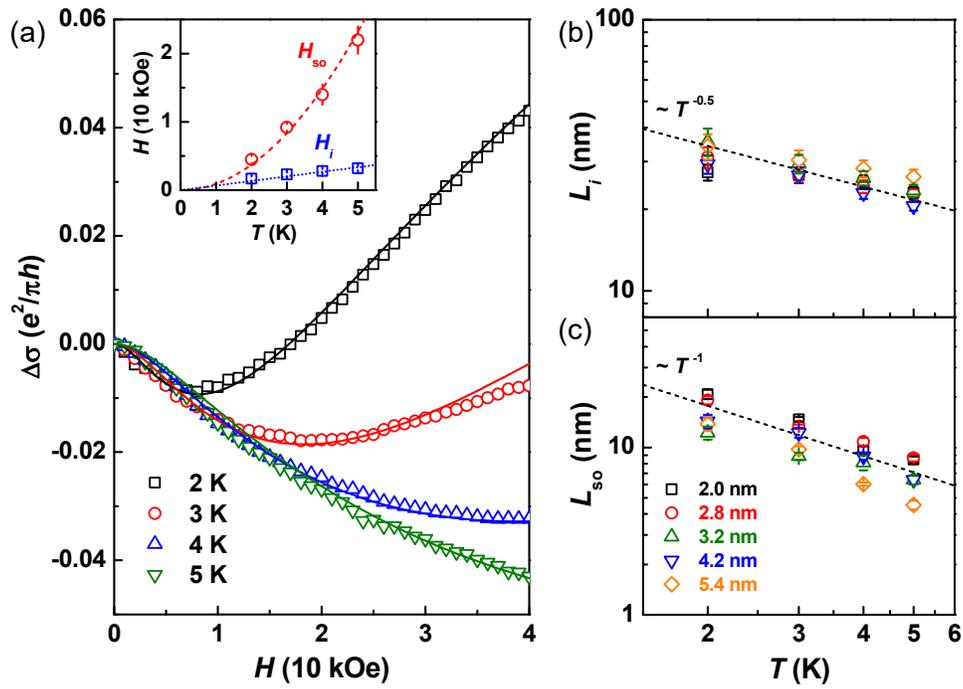

Fig. 3. (a) MC of 2.8 nm SIO at various temperatures with fits to Eq. 4 (solid lines). Inset: $H_i$ and $H_{so}$ vs. $T$ with fits to $T$ (dashed line) and $T^2$ (dotted line) dependences, respectively. (b) $L_i$ and (c) $L_{so}$ vs. $T$ taken on films of various thicknesses. The dashed lines serve as the guide to the eye.

Figure 4

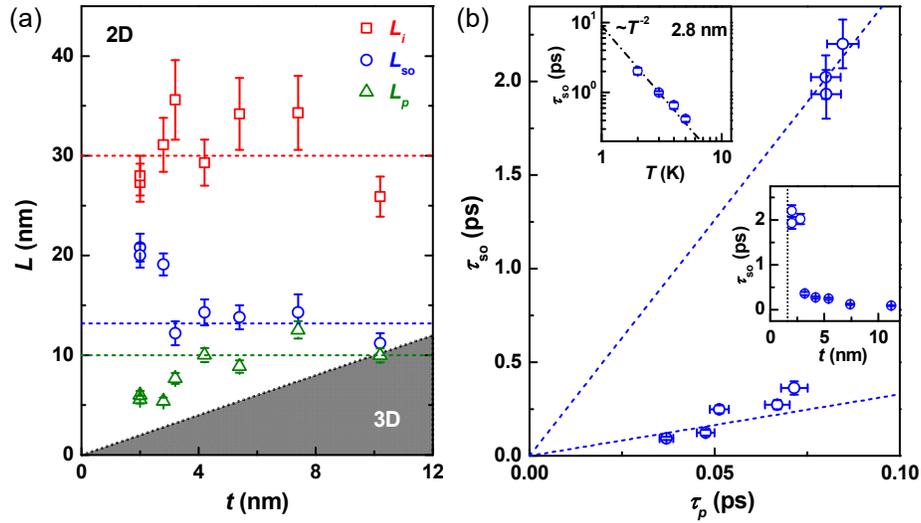

Fig. 4. (a) $L_i$, $L_{so}$ and $L_p$ vs. film thickness. The dotted line set at $L = t$ divides the 2D and 3D (shadowed area) regimes. The dashed lines serve as the guide to the eye. (b) $\tau_{so}$ at 2 K as a function of $\tau_p$ for electrons. The dashed lines are linear fits. Upper inset: $\tau_{so}$ vs. $T$ for the 2.8 nm SIO with a fit to $T^{-2}$-dependence (dash-dot line). Lower inset: $\tau_{so}$ vs. $t$. The dotted line marks the critical thickness for insulating behavior (1.6 nm).